\documentclass{article}

\usepackage{arxiv}

\usepackage[utf8]{inputenc} 
\usepackage[T1]{fontenc}    
\usepackage{hyperref}       
\usepackage{url}            
\usepackage{booktabs}       
\usepackage{amsmath,amssymb,amsfonts}      
\usepackage{dsfont} 
\usepackage{nicefrac}       
\usepackage{microtype}      
\usepackage{lipsum}
\usepackage{graphicx}
\usepackage{subcaption}
\graphicspath{ {./images/} }

\usepackage{enumitem}
\setlist[enumerate]{label=\roman*}
\newtheorem{theorem}{Theorem}

\title{Portfolio Optimization Using a Consistent Vector-Based MSE Estimation Approach}

\author{
Maaz~Mahadi \\
  Department of Electrical and Computer Engineering\\
  King Abdulaziz University\\
  Jeddah, KSA \\
  \texttt{mmahadi@stu.kau.edu.sa} \\
   \And
 Tarig~Ballal \\
  CEMSE\\
  King Abdullah University of Science and Technology (KAUST)\\
  Thuwal, KSA\\
  \And
Muhammad~Moinuddin\\
Department of Electrical and Computer Engineering\\
King Abdulaziz University\\
Jeddah, KSA \\
   \And
Tareq~Y.~Al-Naffouri \\
CEMSE\\
King Abdullah University of Science and Technology (KAUST)\\
Thuwal, KSA\\
  \And
Ubaid~Al-Saggaf\\
Department of Electrical and Computer Engineering\\
King Abdulaziz University\\
Jeddah, KSA \\
}

\long\def\comment#1{}

\DeclareMathOperator*{\argmin}{arg\,min}

\newfont{\bbb}{msbm10 scaled 700}

\newfont{\bb}{msbm10 scaled 1100}

\newcommand{\wv}{{\bf w}}

\newcommand{\xv}{{\bf x}}
\newcommand{\yv}{{\bf y}}

\newcommand{\Bm}{{\bf B}}

\newcommand{\Id}{{\bf I}}

\newcommand{\Tm}{{\bf T}}
\newcommand{\Um}{{\bf U}}

\newcommand{\Xm}{{\bf X}}
\newcommand{\Ym}{{\bf Y}}
\newcommand{\Zm}{{\bf Z}}

\newcommand{\deltav}{\hbox{\boldmath$\delta$}}

\newcommand{\muv}{\hbox{\boldmath$\mu$}}

\newcommand{\Deltam}{\hbox{\boldmath$\Delta$}}
\newcommand{\Sigmam}{\hbox{\boldmath$\Sigma$}}

\newcommand{\Thetam}{\hbox{\boldmath$\Theta$}}

\newcommand{\diag}{{\hbox{diag}}}

\newcommand{\trace}{{\hbox{tr}}}

\begin{document}
\maketitle
\begin{abstract}
This paper is concerned with optimizing the global minimum-variance portfolio's (GMVP) weights in high-dimensional settings where both  observation and  population dimensions grow at a bounded ratio. Optimizing the GMVP weights is highly influenced by the data covariance matrix estimation. In a high-dimensional setting, it is well known that the sample covariance matrix is not a proper estimator of the true covariance matrix since it is not invertible when we have fewer observations  than the data dimension. Even with more observations, the sample covariance matrix may not be well-conditioned. This paper determines the GMVP weights based on a regularized covariance matrix estimator to overcome the aforementioned difficulties. Unlike other methods, the proper selection of the regularization parameter is achieved by minimizing the mean-squared error of an estimate of the noise vector that accounts for the uncertainty in the data mean estimation. Using random-matrix-theory tools, we derive a consistent estimator of the achievable mean-squared error that allows us to find the optimal regularization parameter using a simple line search. Simulation results demonstrate the effectiveness of the proposed method when  the data dimension is larger than the number of data samples or of the same order.
\end{abstract}


\section{Introduction}
Decision making with regards to investment in the stock market has become increasingly more complex because of the dynamic nature of the stocks available to investors and the advent of new unconventional and risky options \cite{2019-bodnar-tests}. Throughout the years, the portfolio optimization problem has attracted the attention of many signal-processing researchers due to its close relationship to the field. The portfolio optimization problem aims at achieving the maximum possible returns with the least volatility percentage \cite{2020-huni-application}. The Economist, Harry Markowitz, introduced the modern portfolio theory, or mean-variance analysis (MVP), in \cite{1952-markowitz-mvp}. Other portfolios such as the global MVP and the maximum sharp ratio portfolio (MSRP) have been proposed as improvements of the MVP. Portfolio optimization utilizes the available financial data to reach conclusions regarding the allocation of wealth to each of the available stocks. The most important measurement in portfolio optimization is the data covariance matrix (CM).

CM estimation in the classical signal processing framework relies on asymptotic statistics of a number of observations, $n$, which is assumed to grow largely compared to the population dimension, $p$, i.e., $n/p \rightarrow \infty$ as $n\rightarrow \infty$ \cite{2012-couillet-signal}. However, many practical applications, such as finance, bioinformatics and data classification, require an estimate of the CM when the data dimension is large compared to the sample size \cite{2019-ollila-optimal}. In such cases, it is well known that the default estimator, i.e., the empirical \emph{sample covariance matrix} (SCM), is usually ill-conditioned, leading to poor performance.

If the case where $p>n$, the SCM is not invertible; whereas for $p<n$, the SCM is invertible but might be ill-conditioned, which substantially increases estimation error. In other words, for a large $p$, it is not practically guaranteed that the number of observations is sufficient to develop a well-conditioned CM estimator \cite{2004-ledoit-well}. Such scenarios have motivated researchers to look into estimation problems in the high-dimensional regime \cite{2012-couillet-signal}.

In scenarios with limited data, a regularized SCM (RSCM) estimator of the following general form is widely used \cite{2019-ollila-optimal}:
\begin{equation}
	\label{eqn:Sigma_beta_gamma}
	\widehat{\Sigmam}_{\beta,\gamma} = \beta \widehat{\Sigmam} + \gamma {\bf I},
\end{equation}
where $\widehat{\Sigmam}$ is the SCM defined in \eqref{eqn:Sigma_hat} (further ahead), $\beta$, $\gamma \in \mathbb{R}^{+}$ are the regularization, or \emph{shrinkage}, parameters. These parameters can be determined based on minimizing the \emph{mean-squared error} (MSE), which results in oracle shrinkage parameters, $\beta_\text{o}$ and $\gamma_\text{o}$, as follows \cite{2019-ollila-optimal, 2004-ledoit-well}:
\begin{equation}
	\label{eqn:oracle_shrinkage_parameters}
	(\beta_\text{o}, \gamma_\text{o}) = \argmin_{\beta, \gamma > 0} \mathbb{E} \left[  \left \| \widehat{\Sigmam}_{\beta,\gamma} -  \Sigmam  \right \|_{\text{F}}^2 \right],
\end{equation}
where $\|.\|_{\text{F}}$ denotes the Forbenius matrix norm. The estimation of $(\beta_\text{o}, \gamma_\text{o})$ based on \eqref{eqn:oracle_shrinkage_parameters} depends on the true CM, $\Sigmam$. To circumvent this issue, Ledoit and Wolf \cite{2004-ledoit-well} proposed a distribution-free consistent estimator of $(\beta_\text{o}, \gamma_\text{o}) $ in high-dimensional settings. The work in \cite{2019-ollila-optimal} assumes that the observations are from unspecified elliptically symmetric distribution.
The consistent estimator proposed in \cite{2015-yang-robust} uses a hybrid CM estimator based on the Taylor's M-estimator and Ledoit-Wolf shrinkage estimator, which suits a \emph{global minimum variance portfolio} (GMVP) influenced by outliers. A similar approach based on the M-estimator is proposed in \cite{2020-ollila-shrinking}, considering $n>p$ with fully automated selection of the shrinkage parameters.
The minimum variance portfolio estimator in \cite{2020-cai-high}  is based on certain sparsity assumptions imposed on the  inverse of the CM. The work presented in \cite{2021-ballal-adaptive} proposes a different RSCM estimator by manipulating the expression of the GMVP weights.

In this paper, we propose a single-parameter CM estimator. Instead of minimizing the MSE, as in \eqref{eqn:oracle_shrinkage_parameters}, we minimize the MSE of the estimation of the sample noise vector. We utilize RMT tools to obtain a consistent estimator of this MSE. The value of the regularization parameter $\gamma$ is selected as the one that minimizes the estimated MSE. By choosing to minimize the MSE of the noise vector's estimation, we consider the inaccuracy of estimating the true mean.

\section{Global Minimum Variance Portfolio}
We consider a time series comprising $\yv_1, \yv_2 \cdots, \yv_L$ logarithmic returns of $p$ financial assets over a certain investment period. We assume that the elements of $\yv_t$, ($t=1,2, \cdots, L$) are independent and identically distributed (i.i.d.) and are generated according to the following stochastic model \cite{2012-rubio-performance}:

\begin{equation}
	\label{eqn:y}
	\yv_t = \muv_t + \Sigmam_t ^{\frac{1}{2}} \xv_t,
\end{equation}
where $\muv_t \in \mathbb{R}^{n \times 1}$ and $\Sigmam_t \in \mathbb{R}^{p \times p}$ are the mean and the CM of the asset returns over the investment period, and $\xv_t$ is an i.i.d. random noise vector of zero mean and identity CM. For simplicity, we drop  the subscript $t$ from  $\muv_t$ and $\Sigmam_t$. For the investment period of interest, we define $\wv \in \mathbb{R}^p$ as the asset holdings vector, also known as the weight vector. The GMVP optimally minimizes the portfolio variance under single-period investment horizon, such that the weight vector is normalized by the outstanding wealth \cite{2012-rubio-performance}, i.e.,

\begin{equation}
	\label{eqn:gmvp_problem}
	\min _{\mathbf{w} \in \mathbb{R}^{p}} \wv^T \Sigmam \wv \quad \text { subject to } \mathds{1}_p^T \wv=1,
\end{equation}
where $\mathds{1}_p$ is a column vector of $p$ 1's.
The solution of \eqref{eqn:gmvp_problem} can be obtained by using the Lagrange-multipliers method, which results in the optimum weights \cite{2015-yang-robust}:

\begin{equation}
	\label{eqn:wo}
	\wv_\text{\tiny{GMVP}}=\dfrac{\Sigmam^{-1} \mathds{1}_p}{\mathds{1}_p^T \Sigmam^{-1} \mathds{1}_p}.
\end{equation}
The CM in \eqref{eqn:wo} is unknown and should be estimated. As stated earlier, the SCM estimate does not perform well because it is usually ill-conditioned; hence, we apply the RSCM estimator and \eqref{eqn:wo} becomes

\begin{equation}
	\label{eqn:w_GMVP}
	\widehat{\wv}_\text{\tiny{GMVP}}=\frac{\widehat{\Sigmam}_\text{RSCM}^{-1} \mathds{1}_p}{\mathds{1}_p^T \widehat{\Sigmam}_\text{RSCM}^{-1} \mathds{1}_p},
\end{equation}
where $\widehat{\Sigmam}_\text{RSCM}$ is the RSCM which can take the form of \eqref{eqn:Sigma_beta_gamma}, for example. In the following section, we develop a RSCM estimator method and properly set the value of its regularization parameter.

\section{The proposed Consistent Vector-Based MSE Estimator}
The SCM, $\widehat{\boldsymbol{\Sigma}}$, and the sample mean, $\widehat{\muv}$, can be estimated from the $n$ past return observations as follows:
\begin{equation}
	\label{eqn:Sigma_hat}
	\widehat{\boldsymbol{\Sigma}} = \dfrac{1}{n-1} \sum_{j=1}^{n} (\yv_{t-j} - \widehat{\muv} )(\yv_{t-j} - \widehat{\muv})^T,
\end{equation}

\begin{equation}
	\label{eqn:mu_hat1}
	\widehat{\muv}  = \dfrac{1}{n} \sum_{j=1}^{n} \yv_{t-j}. 
\end{equation}

We notice that computing $\widehat{\Sigmam}$ using \eqref{eqn:Sigma_hat} involves evaluating the sample mean, not the true mean. This can worsen performance, especially for a small number of observations.
Subtracting $\widehat{\muv}$ from both sides of \eqref{eqn:y}, we obtain

\begin{equation}
	\label{eqn:ytilde}
	{\yv}_t - \widehat{\muv} =  (\widehat{{\Sigmam}}^ {\frac{1}{2}} + \Deltam ) \xv_t+  \deltav,
\end{equation}
where  $\deltav \triangleq \muv - \widehat{\muv}$ and $ \Deltam  =  \Sigmam^{\frac{1}{2}}-\widehat{{\Sigmam}}^ {\frac{1}{2}} $. Eq. \eqref{eqn:ytilde} can be viewed as a linear model with bounded uncertainties in both $\Sigmam^\frac{1}{2}$ and $\muv$ \cite{1998-chandrasekaran-SIAMparameter}.
We seek an estimate, $\widehat{\xv}_t$ that performs well for any allowed perturbation $(\Deltam, \deltav)$ by formulating the following min-max problem \cite{1998-chandrasekaran-SIAMparameter}:
\begin{equation}
	\label{eqn:minmax}
	\min _{\widehat{\xv}_t } \max_{\Deltam, \deltav}~  \left[ \| (\widehat{{\Sigmam}}^ {\frac{1}{2}} + \Deltam ) \widehat{\xv}_t - (\yv_t - \widehat{\muv} - \deltav) \|_2  \right].
\end{equation}
A unique solution can exist which  takes the form \cite{1998-chandrasekaran-SIAMparameter} 
\begin{equation}
	\label{eqn:xt_hat}
	\widehat{\xv}_t = (\widehat{\Sigmam} + \gamma \mathbf{I})^{-1} \widehat{\Sigmam}^ {\frac{1}{2}} (	{\yv}_t - \widehat{\muv} ).
\end{equation}

$\widehat{\xv}_t  $ is a function of $\gamma$, which when properly set leads to the best estimate of $\xv_t$. It is easy to recognize that $(\widehat{\Sigmam} + \gamma \mathbf{I})^{-1}$ can be used as an estimator of the CM inverse, i.e., $\widehat{\Sigmam}_\text{RSCM}^{-1} = \widehat{\Sigmam}_\gamma^{-1}= (\widehat{\Sigmam} + \gamma \mathbf{I})^{-1}$. Such estimator is widely used in the literature, e.g.,  \cite{2007-guo-regularized,2015-zollanvari-generalized,2017-ballal-bounded,2020-elkhalil-large,1988-carlson-covariance,2003-li-robust,2021-mahadi-low,2021-mahadi-robust,2018-suliman-robust}; to name a few. The optimal value of $\gamma$ that estimates $\widehat{\Sigmam}_\gamma^{-1}$ is the one that minimizes the MSE  for estimating  ${\xv}_t$. That is
\begin{align}
	\text{MSE}(\gamma) &= \mathbb{E} \left[ \| \xv_t - \widehat{\xv}_t \|_2^2 \right] \\
	& = \mathbb{E} \left[ \| \xv_t - \widehat{\Sigmam}_\gamma ^{-1}\widehat{\boldsymbol{\Sigma}}^ {\frac{1}{2}} ( {\Sigmam}^ {\frac{1}{2}} \xv_t+  \deltav)      \|_2^2 \right]. \label{eqn:MSE}
\end{align} 
We choose the optimal $\gamma_\text{o}$ as follows:
\begin{equation}
	\gamma_\text{o}  = \argmin \text{MSE}(\gamma).
\end{equation}
The choice of minimizing the MSE is reasonable because, under certain conditions, the minimization problem in \eqref{eqn:gmvp_problem} and the minimum MSE are equivalent \cite{1993-kay-fundamentals}, \cite{2008-rubio-generalized}.
Unlike the other methods, it is remarkable that the uncertainty in estimating the mean is incorporated in \eqref{eqn:MSE}. We expect the effect of the uncertainty in the mean estimation to be high when we have a limited number observations. Also, unlike the methods that are based on \eqref{eqn:oracle_shrinkage_parameters},  when we search for the optimal $\gamma$ that minimizes \eqref{eqn:MSE}, we actually estimate the inverse of the CM rather than estimating the CM itself. This is important because we use it in \eqref{eqn:w_GMVP}.
We obtain the following normalized (by $n$) expression of the MSE (see Appendix~\ref{sec:appB}):
\begin{equation}\label{eqn:MSE1}
	\begin{split}
		\text{MSE} 
		(\gamma) 
		= \frac{p}{n} + 	\dfrac{n+1}{n}\mathbb{E} \Bigl[ \underbrace{ \dfrac{1}{n}\trace \bigl[ \Sigmam \widehat{\Sigmam} \bigl( \widehat{\Sigmam}  + \gamma {\bf I}   \bigr)^{-2}  \bigr]}_{A(\gamma)} \Bigr] \\
		-2 \mathbb{E} \bigl[  \underbrace{ \dfrac{1}{n}\trace \bigl[ \Sigmam^\frac{1}{2} \widehat{\Sigmam}^\frac{1}{2} \bigl( \widehat{\Sigmam}  + \gamma {\bf I}   \bigr)^{-1}  \bigr] }_{B(\gamma)}  \bigr].
	\end{split}
\end{equation}

We observe that \eqref{eqn:MSE1} is expressed in terms of the unknown quantity, $\boldsymbol{\Sigma}$. In this case, using a direct plugin formula, i.e., substituting $\Sigmam$ with $\widehat{\Sigmam}$ results in 

\begin{equation}\label{eqn:MSE_Pluging}
	\begin{split}
		\widehat{\text{MSE}}_{\text{plugin}}
		(\gamma) 
		= \frac{p}{n} + \dfrac{n+1}{n}\mathbb{E} \Bigl[  \dfrac{1}{n}\trace \bigl[ \widehat{\Sigmam}^2 \bigl( \widehat{\Sigmam}  + \gamma {\bf I}   \bigr)^{-2}  \bigr] \Bigr] \\
		-2 \mathbb{E} \bigl[  \dfrac{1}{n}\trace \bigl[  \widehat{\Sigmam} \bigl( \widehat{\Sigmam}  + \gamma {\bf I}   \bigr)^{-1}  \bigr]  \bigr].
	\end{split}
\end{equation}

\begin{figure}
	\centering
	\includegraphics[width=\linewidth]{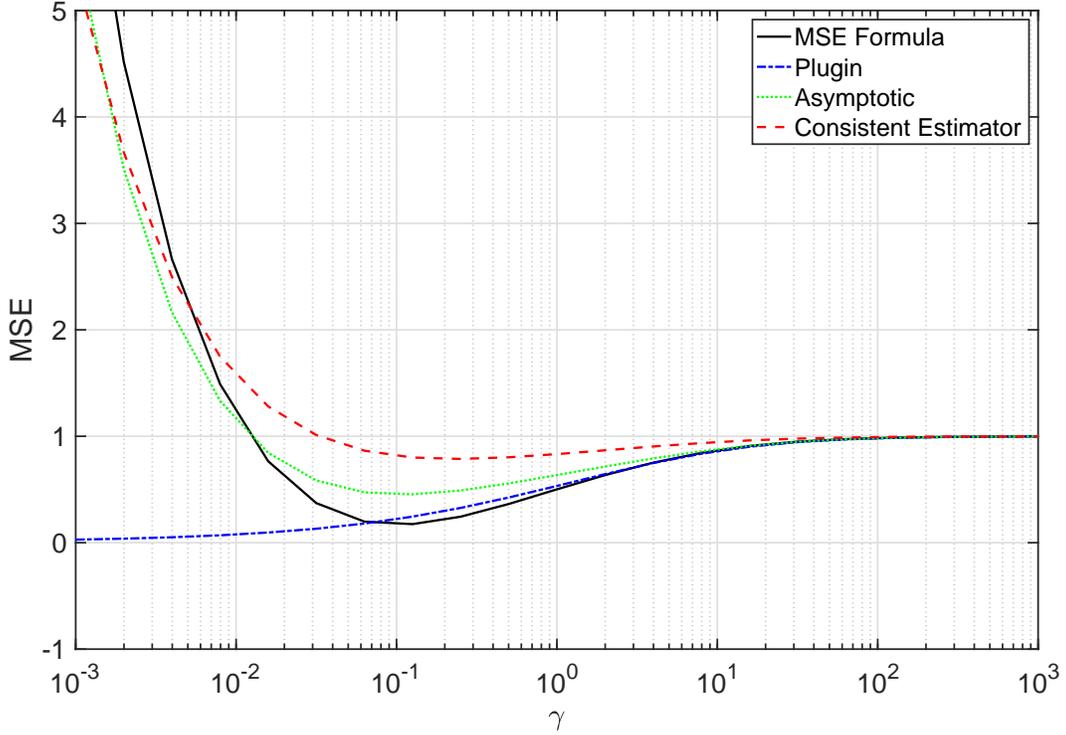}
	\caption{Different MSE curves versus the regularization parameter: the derived formula of the MSE, Eq.~\eqref{eqn:MSE1}, the plugin estimator, Eq.~\eqref{eqn:MSE_Pluging}, the asymptotic curve, Eq~\eqref{eqn:MSE_DE}, and the consistent MSE Eq.~\eqref{eqn:MSE_consistent}. The results are generated from Gaussian data that follows \eqref{eqn:y} with $p = n =300,$ $[\Sigmam]_{i,j} = 0.6^{|i-j| }$ and $\muv = \mathds{1}_p$.}
	\label{fig:MSE}	
\end{figure}

However, the estimator in \eqref{eqn:MSE_Pluging} is an inconsistent estimator in the regime where $n$ and $p$ grow at constant rate \cite{2020-lama-asymptotic}. To clarify, Fig.~\ref{fig:MSE}  plots an example of the derived MSE$(\gamma)$\eqref {eqn:MSE1} and the plugin estimation method \eqref{eqn:MSE_Pluging} versus a wide range values of $\gamma$. It is clear that using the plugin strategy does not help obtain the minimum MSE suitably. Instead, as the figure depicts, the plugin estimation method selects an improper $\gamma$ that corresponds to a high MSE.

As an alternative remedy , we seek a consistent estimator of \eqref{eqn:MSE1} by leveraging tools from RMT. To this end, we need to first obtain an asymptotic expression of \eqref{eqn:MSE1}. To do so, the following assumption should hold true.

\textit{Assumption 1}: As $p$,$n$ $\rightarrow \infty$, $p/n \rightarrow c\in (0,\infty)$.

Assumption 1 leads to the following theorem:

\begin{theorem}
	Under Assumption 1, MSE($\gamma$) in \eqref{eqn:MSE1}  asymptotically converges to
	\begin{align}
		\label{eqn:MSE_DE}
		&\text{MSE} (\gamma) \asymp \frac{p}{n} +\dfrac{n+1}{n}. \dfrac{1}{n} (\tilde{\delta}_1 + \gamma \tilde{\delta}_1^\prime) \trace \Bigl[ \Sigmam^2\bigl( \tilde{\delta}_1\Sigmam + \gamma {\bf I} \bigr)^{-2}   \Bigr] \nonumber\\
		&-  \dfrac{2}{n} \Re \Bigg[ \trace \Bigl[  \Sigmam^\frac{1}{2} \bigl( \tilde{\delta}_2\Sigmam^\frac{1}{2} -i \sqrt{\gamma} {\bf I} \bigr)^{-1} \Bigr] \Bigg],
	\end{align}
	where $\tilde{\delta}_1$ is the unique positive solution to the following system of equations:
	
	\begin{equation}
		\label{eqn:delta_1}
		\left\{\begin{array}{l} 
			\delta_1 = \frac{1}{n} \trace \bigl[ \Sigmam \bigl( \tilde{\delta}_1 \Sigmam + \gamma {\bf I}_p \bigr)^{-1}   \bigr],  \\
			\tilde{\delta}_1 = \frac{1}{n} \trace \bigl[ \Tm \bigl( \delta_1 \Tm + {\bf I}_n \bigr)^{-1}  \bigr],
		\end{array}\right.
	\end{equation}
	where $\Tm = \diag ([1,1, \cdots,1,0]^T) \in \mathbb{R}^{n \times n}$; hence, $\tilde{\delta}_1$ can be written as follows:
	\begin{equation}
		\label{eqn:tilde_delta_1}
		\tilde{\delta}_1 = \dfrac{1}{1+\delta_1}.
	\end{equation}
	Similarly, ${\tilde{\delta}}_2$ is obtained by solving
	\begin{equation} 
		\label{eqn:delta_2}
		\left\{\begin{array}{l}
			\delta_2 = \frac{1}{n} \trace \bigl[ \Sigmam^\frac{1}{2} \bigl( \tilde{\delta}_2 \Sigmam^\frac{1}{2} -i \sqrt{\gamma} {\bf I}_p \bigr)^{-1}   \bigr],  \\
			\tilde{\delta}_2 = \frac{1}{n} \trace \bigl[ \Tm \bigl( \delta_2 \Tm + {\bf I}_n \bigr)^{-1}  \bigr].
		\end{array}\right.	
	\end{equation}
	and 
	\begin{equation}
		\label{eqn:tilde_delta_2}
		\tilde{\delta}_2 = \dfrac{1}{1+\delta_2}.
	\end{equation}
\end{theorem}

\textit{Proof:} see  Appendix \ref{sec:appC}.

Now, we are in a position to reveal the consistent estimator of \eqref{eqn:MSE1}.

\begin{theorem}
	Under Assumption 1, the consistent estimator of \eqref{eqn:MSE1} is given by \eqref{eqn:MSE_consistent}
	\begin{align}
		\widehat{\text{MSE}} (\gamma) &\asymp  \frac{p}{n} -2 \Re(\hat{\delta}_2) +\frac{n+1}{n}\frac{(1+\hat{\delta}_1)^2}{\hat{\delta}_1^\prime } .\nonumber \\
		&\Bigl[ (\hat{\tilde{\delta}}_1+ \gamma\hat{\tilde{\delta}}_1^\prime)\hat{\delta}^\prime_1 - \dfrac{1}{n} \trace \bigl[  \widehat{\Sigmam} \bigl(  \widehat{\Sigmam} + \gamma {\bf I}_p \bigr)^{-2}  \bigr]  \Bigr],   \label{eqn:MSE_consistent}
	\end{align}
	where  $\hat{\delta}_1$ and $\hat{\delta}_2$  are the consistent estimators of $\delta_1$ and $\delta_2$, respectively,  and are given by 
	\begin{align}
		\hat{\delta}_1 &= \dfrac{ \frac{1}{n}\trace \bigl[ \widehat{\Sigmam} \bigl( \widehat{\Sigmam}  + {\gamma} {\bf I}_p \bigr)^{-1} \bigr] }{1-\frac{1}{n}\trace \bigl[ \widehat{\Sigmam} \bigl( \widehat{\Sigmam}  + {\gamma} {\bf I}_p \bigr)^{-1} \bigr] }, \label{eqn:delta_1_hat}\\
		\hat{\delta}_2 &=\dfrac{ \frac{1}{n}\trace \bigl[ \widehat{\Sigmam}^{\frac{1}{2}} \bigl( \widehat{\Sigmam}^{\frac{1}{2}}  - i\sqrt{\gamma} {\bf I}_p \bigr)^{-1} \bigr] }{1-\frac{1}{n}\trace \bigl[ \widehat{\Sigmam}^{\frac{1}{2}} \bigl( \widehat{\Sigmam}^{\frac{1}{2}}  - i\sqrt{\gamma} {\bf I}_p \bigr)^{-1} \bigr] }. \label{eqn:delta_2_hat}
	\end{align}
\end{theorem}
\textit{Proof:} see Appendix \ref{sec:appD}.

Back to Fig.~\ref{fig:MSE} which compares the derived MSE with the asymptotic formula \eqref{eqn:MSE_DE} and the consistently estimated MSE \eqref{eqn:MSE_consistent}. It can be seen clearly that the consistent MSE is more suitable to obtain the value of $\gamma$ that minimizes \eqref{eqn:MSE1}. 

A closed form solution for $\gamma$ in \eqref{eqn:MSE_consistent} is infeasible, so we rely on using a line search, where we search for $\gamma$ that minimizes \eqref{eqn:MSE_consistent} within a predefined range.    

\subsection{Summary of the proposed VB-MSE (vector based-MSE) method for Portfolio Optimization}
\begin{enumerate}[start=1,label={\bfseries ~\arabic*.},leftmargin=*,labelindent=0em]
	\item From the historical data estimate $\widehat{\Sigmam}$  using Eq. \eqref{eqn:Sigma_hat}. 
	\item Find the  regularization parameter value, $\gamma_\text{o}$,  that minimizes $\widehat{\text{MSE}} (\gamma)$ in \eqref{eqn:MSE_consistent} using a line search.
	\item Use  $\gamma_\text{o}$ to compute $\widehat{\Sigmam}_{\gamma_\text{o}}^{-1}= (\widehat{\Sigmam} + \gamma_\text{o} \mathbf{I})^{-1}$.
	\item Calculate $\widehat{\wv}_\text{\tiny{GMVP}}$ from \eqref{eqn:w_GMVP} by using $\widehat{\Sigmam}_\text{RSCM}^{-1} = \widehat{\Sigmam}_{\gamma_\text{o}}^{-1}$.
\end{enumerate}
\begin{figure*}[!h]
	\centering
	\begin{subfigure}[t]{0.49\linewidth}
		\centering
		\includegraphics[width=\linewidth]{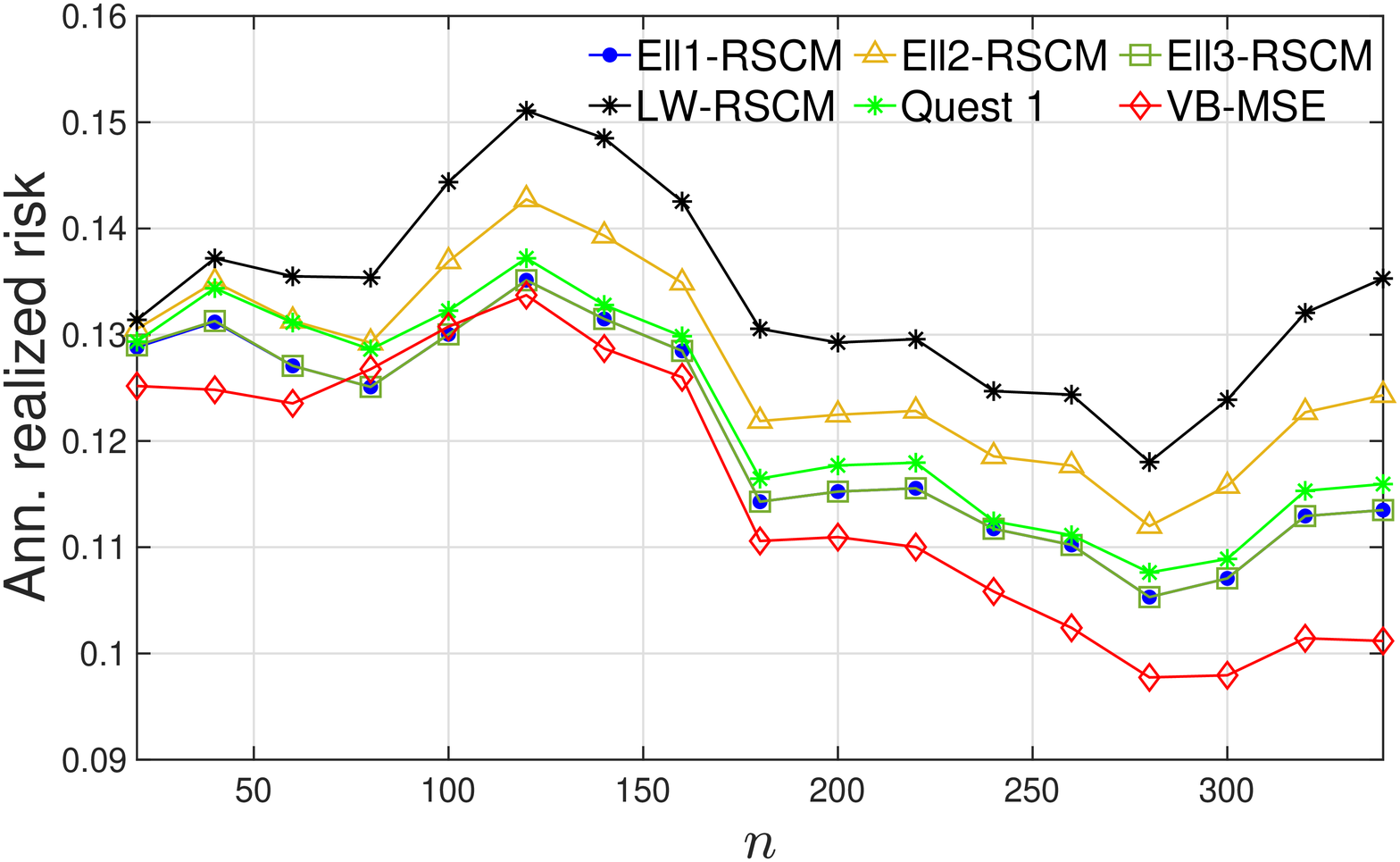}%
		\caption{\centering S\&P 100,  2 Jan. 15 -- 30 Dec. 16, $(p=97)$.}
		\label{fig:SP11-12}		
	\end{subfigure}
	\hfil
	\begin{subfigure}[t]{0.49\linewidth}
		\includegraphics[width=\linewidth]{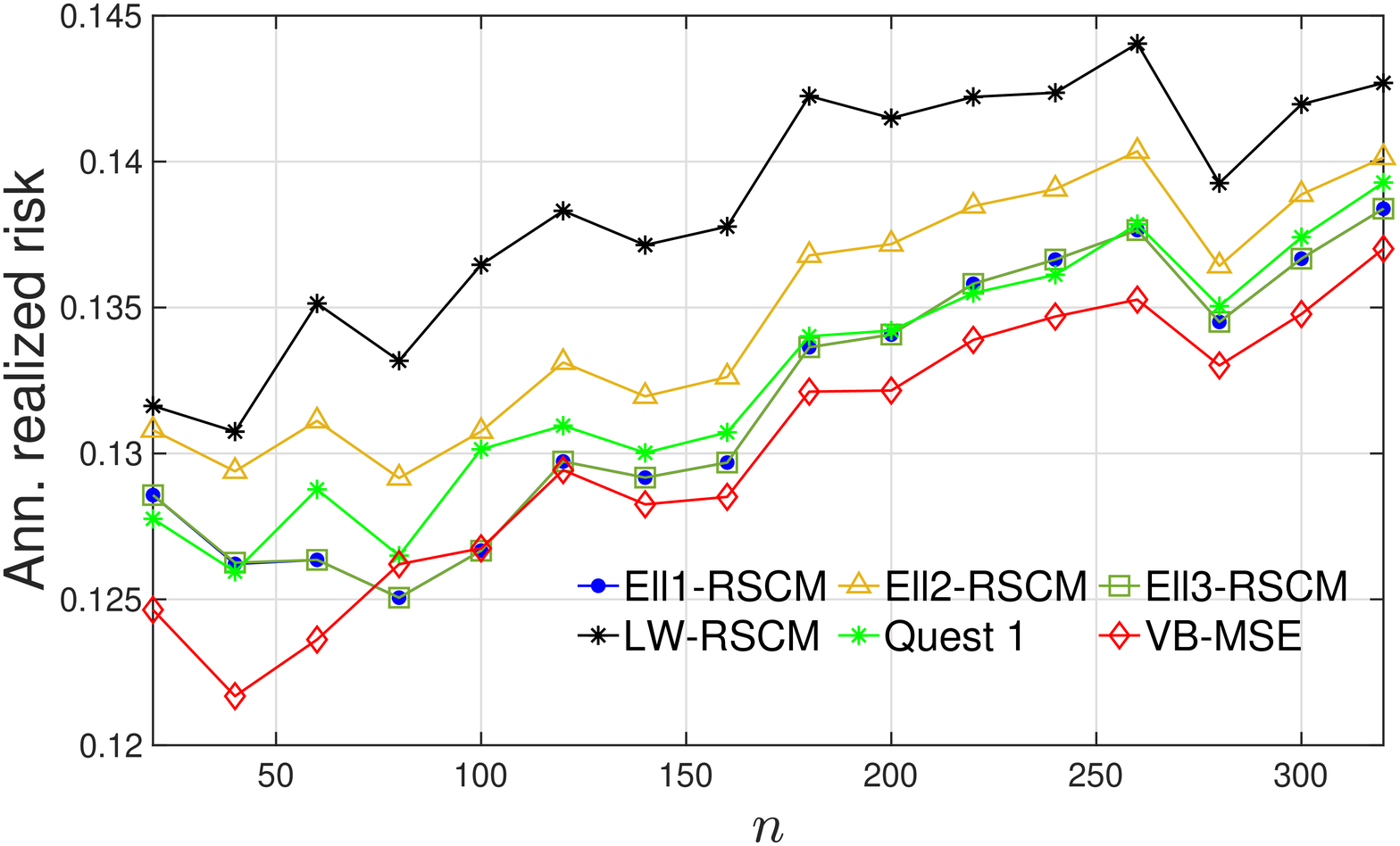}%
		\caption{\centering S\&P 100, 7 Jan. 14 -- 31 Dec. 15, $(p=97)$.}
		\label{fig:SP15-16}		
	\end{subfigure}	
	\begin{subfigure}[t]{0.49\linewidth}
		\includegraphics[width=\linewidth]{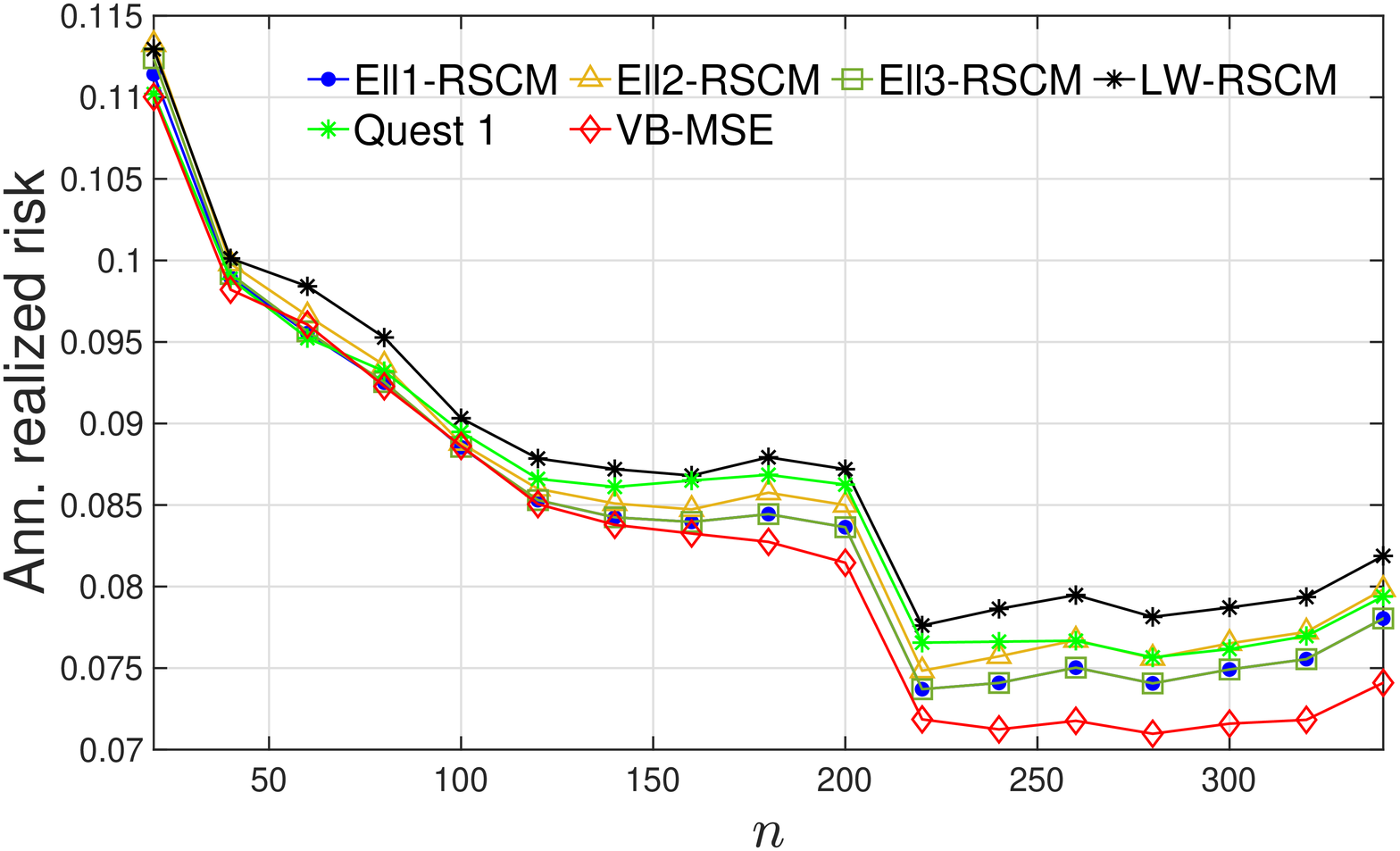}%
		\caption{\centering HSI, 1 Jan. 16 -- 27 Dec. 17, $(p=50)$.}
		\label{fig:HSI14-15}	
	\end{subfigure}	
	\hfil
	\begin{subfigure}[t]{0.49\linewidth}
		\includegraphics[width=\linewidth]{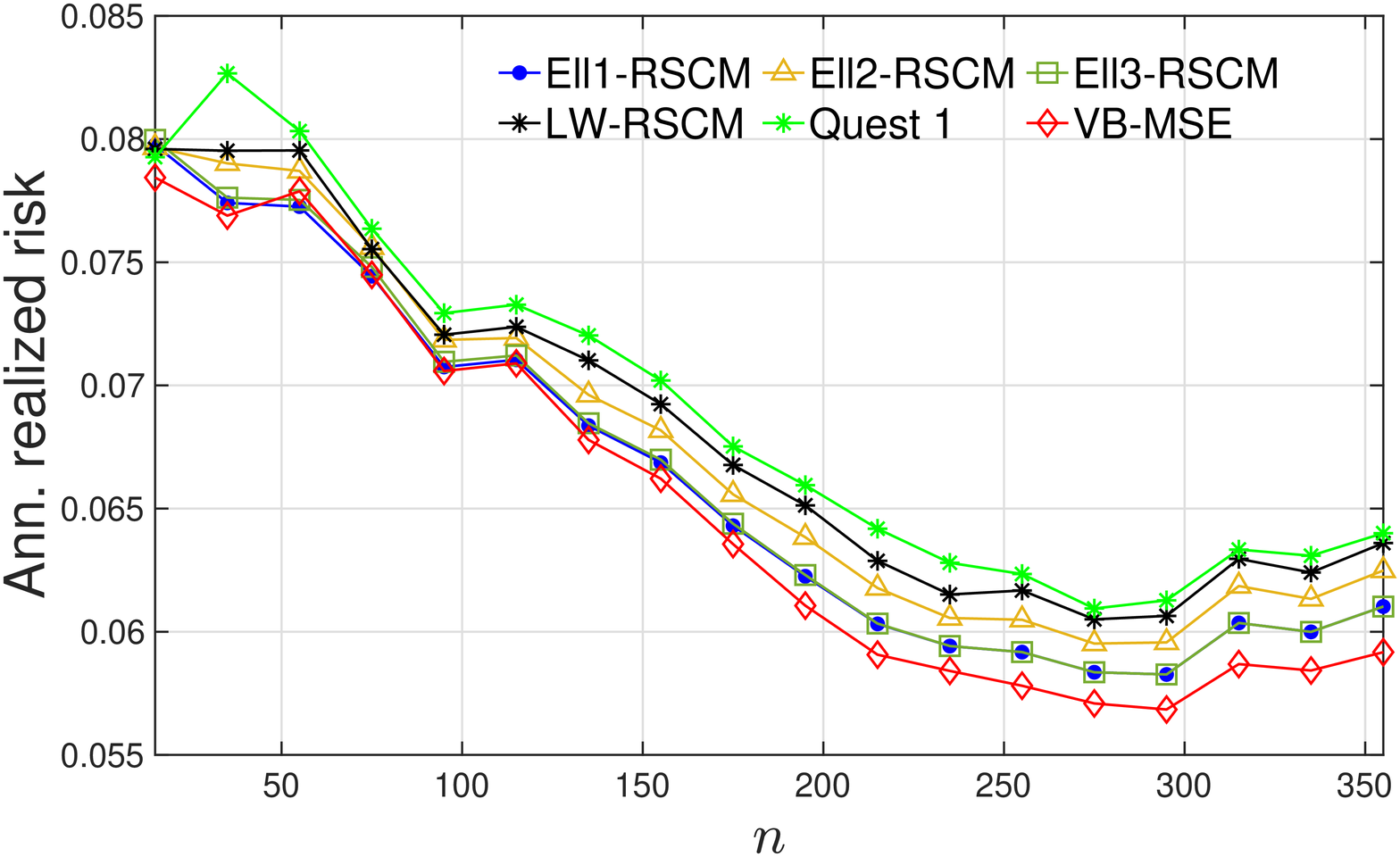}%
		\caption{\centering XMI, 4 Jan. 16 -- 29 Dec. 17, $(p=19)$.}
		\label{fig:XMI16-17}	
	\end{subfigure}	
	\begin{subfigure}[t]{0.49\linewidth}
		\includegraphics[width=\linewidth]{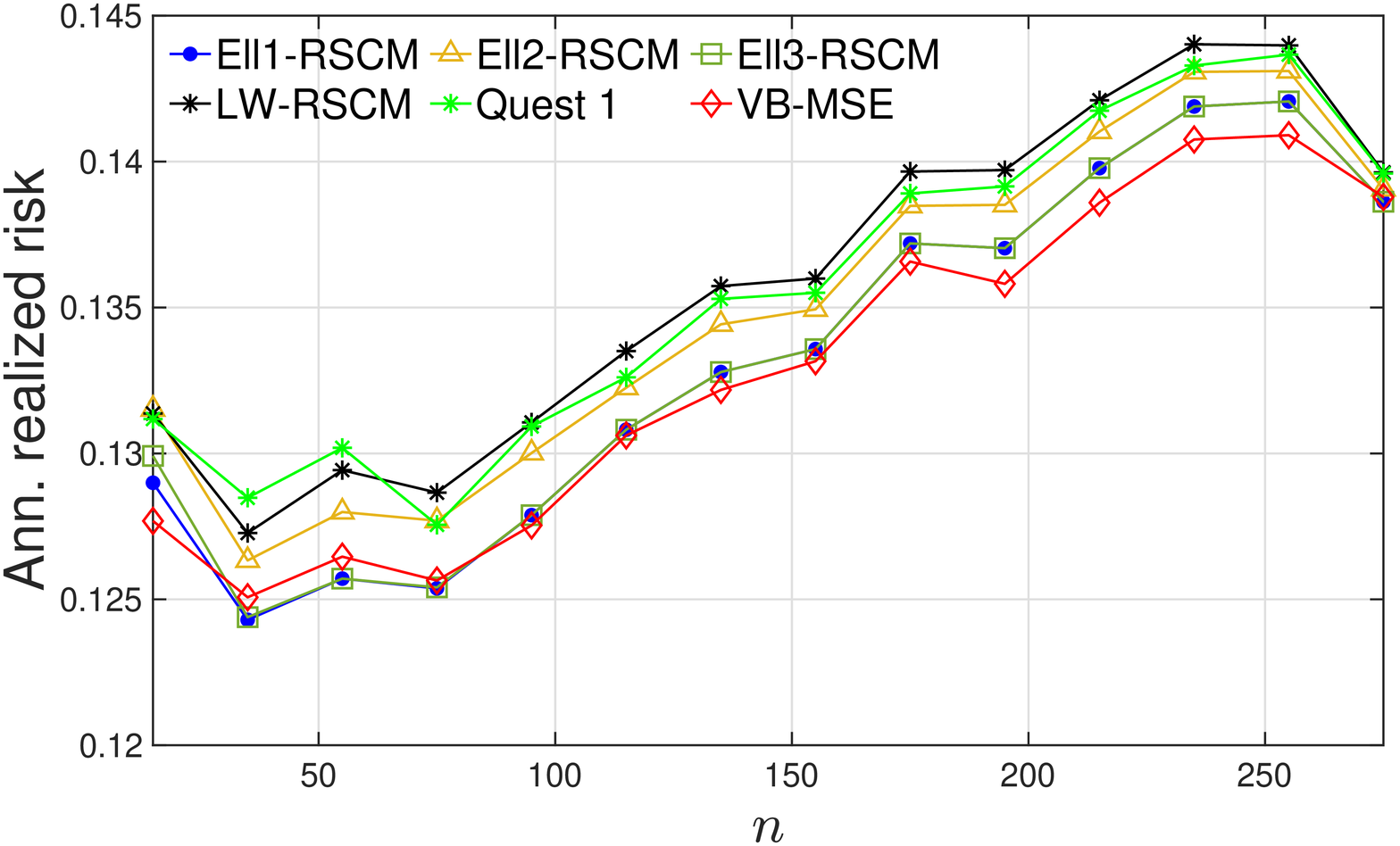}%
		\caption{\centering XMI, 10 Jan. 14 -- 31 Dec. 15, $(p=19)$.}
		\label{fig:XMI14-15}	
	\end{subfigure}	
	\hfil
	\begin{subfigure}[t]{0.49\linewidth}
		\includegraphics[width=\linewidth]{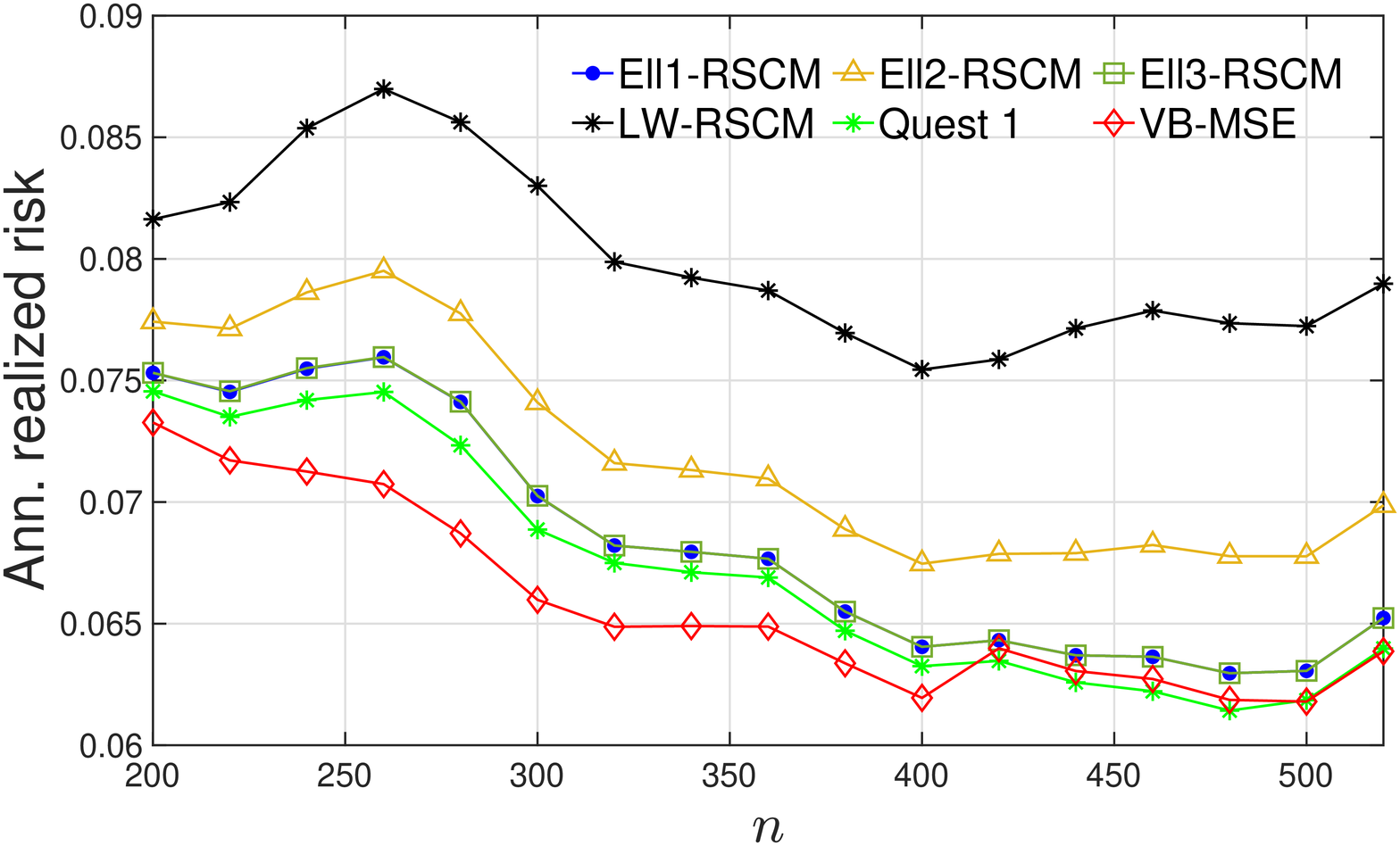}%
		\caption{\centering S\&P 500, 7 Jan. 15 -- 22 Dec. 17, $(p=484)$.}
		\label{fig:SP500_15-17}	
	\end{subfigure}		
	\caption{Annualized realized risk versus training window length for different stock indices.}
	\label{fig:StockInxs}
\end{figure*}

\section{Performance Evaluation}
As conventionally described in the financial literature, we implement the \emph{out-of-sample} strategy defined in terms of a rolling window method (see \cite{2015-yang-robust}). 
At a particular day $t$, the training window for CM estimation is formed from the previous $n$ days, i.e., from $t-n$ to $t-1$, to design the portfolio weights, $\widehat{\wv}_{\text{GMVP}}$. The portfolio returns in the following 20 days are computed based on these weights. Next, the window is shifted 20 days forward and the returns for another 20 days are computed. The same procedure is repeated until the end of the data. Finally, the realized risk is computed as the standard deviation of the returns.
The following list describes the data from different stock market  indices used in our evaluation:
\begin{itemize}
	\item \textbf{Standard and Poor’s 500 (S\&P 500) index}: This index includes 500 companies. The net returns of 484 stocks $(p = 484)$ are obtained for 784 working days between 7 Jan. 2015 and 22 Dec. 2017.
	\item \textbf{Standard and Poor’s 100 (S\&P 100) index}: The index is a subset of the S\&P 500 that comprises 100 stocks.
	We consider two different periods to obtain the net returns from different stocks \cite{sp100}. The first period is from 7 Jan. 2014 to 31 Dec. 2015 (501 trading days), where we fetch data of 97 stocks $(p=97)$. The second period is from 2 Jan., 2015 to 30 Dec. 2016 (504 trading days) that contains net returns of 97 stocks $(p=97)$.
	\item \textbf{NYSE Arca Major Market Index (XMI)}: This market index is made up of 20 Blue Chip industrial stocks of major U.S. corporations \cite{xmi}. A full length time series contains 503 working days from 4 Jan., 2016 to 29 Dec. 2017 is obtained for 19 stocks $(p=19)$. The second period is from 10 Jan. 2014 to 31 Dec. 2015 (498 working days).
	\item \textbf{Hang Seng Index (HSI)}: This market index comprises 50 stocks \cite{hsi}. The returns of all the stocks $(p=50)$ is obtained from 1 Jan. 2016 to 27 Dec. 2017 (491 trading days).
\end{itemize}

Fig.~\ref{fig:StockInxs} shows the  annualized realized risk of the aforementioned market indices versus the number of training samples. We compare the proposed vector-based method, VB-MSE, against the elliptical estimators ELL1-RSCM, ELL2-RSCM and ELL3-RSCM  \cite{2019-ollila-optimal}, \cite{2018-ollila-matlab}, the Ledoit-Wolf estimator, LW-RSCM \cite{2004-ledoit-well}, \cite{2018-ollila-matlab}, the nonlinear estimator Quest 1 \cite{2017-ledoit-numerical} .

Fig.~\ref{fig:StockInxs}~(\subref{fig:SP11-12}) plots the result of the S\&P 100 index from 2 Jan. 2015 to 30 Dec. 2016. As can be seen from the figure, the performance of the proposed VB-MSE method outperforms all other the methods except at $n= 80$ and  $100$, where it  is slightly worse than Ell1-RSCM aand Ell3-RSCM. Similarly, VB-MSE has a superior performance in Fig.~\ref{fig:StockInxs}~(\subref{fig:SP15-16}), which plots the result from 7 Jan. 2014 to 31 Dec. 2015. However, at $n = 20$ and $80$ Ell1-RSCM and Ell3-RSCM perform better. The realized risk for the HSI index is depicted in Fig.~\ref{fig:StockInxs}~(\subref{fig:HSI14-15}) from 1 Jan. 2016 -- 27 Dec. 2017. The proposed method has a comparable performance to Quest 1, Ell1-RSCM and Ell3-RSCM at $n = 20, 40$ and $60$  but it outperforms all the methods for $ 100 <n\le 340$. The results of the XMI index from 4 Jan. 2016 -- 29 Dec. 2017 and from 10 Jan. 2014 -- 31 Dec. 2015 are shown in Fig.~\ref{fig:StockInxs}~(\subref{fig:XMI16-17}) and Fig.~\ref{fig:StockInxs}~(\subref{fig:XMI14-15}), respectively. Overall, in both figures, VB-MSE is the best performing method.
Finally, Fig.~\ref{fig:StockInxs}~(\subref{fig:SP500_15-17}) plots the realized risk of the S\&P 500 index from 10 Jan. 2015 -- 31 Dec. 2017. The figure shows clearly that the proposed method outperforms the other methods when $  200 \le n \le 400$. 

From Fig.~\ref{fig:StockInxs}~(\subref{fig:SP11-12}) -- (\subref{fig:SP500_15-17}), we can conclude that, on average, the proposed VB-MSE method compares favorably to all the benchmark methods tested in this paper. The method is also more consistent over the various datasets.

\section{Conclusion}
In this paper, we have proposed a regularized covariance matrix estimator under high-dimensionality settings. The proposed method searches for the optimal regularization parameter based on a consistent estimator of the MSE of the estimated  vector. Portfolio optimization results from real financial data show that the proposed method performs reasonably well and outperforms a host of benchmark methods.

\section{Mathematical Tools}
\label{sec:appA}
For convenience, we write Equation \eqref{eqn:y} in matrix form
\begin{equation}
	\label{eqn:Ycapital}
	\Ym = \Sigmam^{\frac{1}{2}}\Xm + \muv \	\mathds{1}_n, 
\end{equation}
where $\Xm = [\xv_1 \xv_2 \cdots \xv_n]$ with $\xv_i \sim \mathcal{N}(\boldsymbol{0}, \Id_p)$. We need to express the SCM in \eqref{eqn:Sigma_hat} in an appropriate matrix form as well, as follows:
\begin{align}
	\widehat{\Sigmam}  = \dfrac{1}{n-1} \Bm\Bm^T,
\end{align}
where $\Bm \in \mathbb{R}^{p \times n}$. It can be immediately recognized from  \eqref{eqn:Sigma_hat} that $\Bm$ is
\begin{equation}
	\label{eqn:B2}
	\Bm = \Ym - \widehat{\muv} \mathds{1}_n^T.
\end{equation}
Also, we can easily verify that
\begin{equation}
	\label{eqn:mu_muhat}
	\widehat{\muv} = \muv + \dfrac{1}{n} \Zm \mathds{1}_n,
\end{equation}
where $\Zm \triangleq \Sigmam^{\frac{1}{2}} \Xm$.  
Finally, we perform the following operations to reach the model of $\widehat{{\Sigmam}}$ at the end: 
\begin{align}
	\label{eqn:S_hat2}
	\widehat{\Sigmam} &= \dfrac{1}{n-1} \left(  \Zm\Zm^T - \Zm \dfrac{\mathds{1}_n\mathds{1}_n^T}{n} \Zm^T  \right) \nonumber\\
	&= \dfrac{1}{n-1}  \Zm\left( \Id_n - \dfrac{\mathds{1}_n\mathds{1}_n^T}{n}  \right) \Zm^T \nonumber\\
	&= \dfrac{1}{n-1}  \Zm\Um\Tm\Um^T \Zm^T\nonumber\\
	&= \dfrac{1}{n-1}  \Sigmam^{\frac{1}{2}}\Xm\Um\Tm\Um^T \Xm^T \Sigmam^{\frac{1}{2}} \\
	&= \dfrac{1}{n-1}  \Sigmam^{\frac{1}{2}}\widetilde{\Xm}\Tm \widetilde{\Xm}^T  \Sigmam^{\frac{1}{2}}, \label{eqn:RMTmodel}
\end{align}
where $\Um$ and $\Tm$ are the matrices of eigenvalue vectors and eigenvalues, respectively, of $( \Id_n - \dfrac{\mathds{1}_n\mathds{1}_n^T}{n})$ obtained using the eigenvalue decomposition. The Gaussian distribution is invariant when multiplying by a unitary matrix; hence, $\widetilde{\Xm}$ has the same distribution as $\Xm$ \cite{2015-zollanvari-generalized}, \cite{2020-elkhalil-large}. 

The model \eqref{eqn:RMTmodel} is a well-established model is the RMT literature. Based on this model, for $z \in \mathbb{C} - \mathbb{R}^+$ and bounded $\Thetam \in \mathbb{R}^{p \times p}$ the following relations, which will be used throughout the derivations, hold true (under Assumption 1)  \cite{2012-rubio-performance}: 
\begin{align}
	&\trace \left[ \Thetam \left( \widehat{\Sigmam} -z \Id_p  \right)^{-1} \right] \asymp \trace \left[ \Thetam \left( \tilde{\delta}\Sigmam -z \Id_p  \right)^{-1} \right] \label{eqn:DE_1} \\
	&\trace \left[ \Thetam  \widehat{\Sigmam}  \left( \widehat{\Sigmam} -z \Id_p  \right)^{-1} \right] \asymp \tilde{\delta}\trace \left[ \Thetam \left( \tilde{\delta}\Sigmam -z \Id_p  \right)^{-1} \right] \label{eqn:DE_11} \\	
	&\trace \left[ \Thetam \left( \widehat{\Sigmam} -z \Id_p  \right)^{-2} \right] \asymp \trace \left[ \Thetam \Sigmam \left( \tilde{\delta}\Sigmam -z \Id_p  \right)^{-2} \right] \label{eqn:DE_2} \\
	&\trace \left[ \Thetam\widehat{\Sigmam} \left( \widehat{\Sigmam} -z \Id_p  \right)^{-2} \right] \asymp (\tilde{\delta} -z\tilde{\delta}^\prime)\trace \left[ \Thetam \Sigmam \left( \tilde{\delta}\Sigmam -z \Id_p  \right)^{-2} \right]. \label{eqn:DE_22} 		
\end{align}

\section{Deriving the MSE formula}
\label{sec:appB}
The MSE in \eqref{eqn:MSE1} can be easily obtained from expanding \eqref{eqn:MSE} and computing the resulted terms. The first term results from  $\mathbb{E} [ \xv_t\xv_t^T] = \Id_p$. The second term is computed as follows:  
\begin{equation}
	\trace \left[\mathbb{E}  \left[ \widehat{\xv}_t\widehat{\xv}^T\right]\right]= \trace \left[ \mathbb{E} \left[ \widehat{\Sigmam}^\frac{1}{2}  \widehat{\Sigmam}_\gamma  ^{-1} \tilde{\yv}\tilde{\yv}^T\widehat{\Sigmam}_\gamma  ^{-1}  \widehat{\Sigmam}^\frac{1}{2}   \right] \right].
\end{equation}
where $\tilde{\yv} \triangleq(\yv - \widehat{\muv})= \left( \Sigmam^\frac{1}{2} \xv_t  + \deltav   \right) $. Using the fact that the expectation and the trace are interchangeable and the cyclic property of traces, we can write
\begin{align}
	\trace \left[\mathbb{E}  \left[ \widehat{\xv}_t\widehat{\xv}_t^T\right]\right]
	&= \trace \left[ \mathbb{E} \left[ \widehat{\Sigmam}_\gamma  ^{-1}   \widehat{\Sigmam}  \widehat{\Sigmam}_\gamma  ^{-1} \left( \Sigmam^\frac{1}{2} \xv_t  + \deltav   \right)\left( \Sigmam^\frac{1}{2} \xv_t  + \deltav   \right)^T  \right] \right].
\end{align}
Also, using the eigenvalue decomposition of $\widehat{\Sigmam} $, it is easy to prove that
\begin{align}
	\widehat{\Sigmam}_\gamma  ^{-1}   \widehat{\Sigmam} &=  \left(  \widehat{\Sigmam}  + \gamma\Id_p   \right)^{-1}\widehat{\Sigmam} \\
	&=  \widehat{\Sigmam}  \left(  \widehat{\Sigmam}  + \gamma\Id_p   \right)^{-1} \\
	&= \widehat{\Sigmam}^\frac{1}{2}   \left(  \widehat{\Sigmam}  + \gamma\Id_p   \right)^{-1}  \widehat{\Sigmam}^\frac{1}{2}.
\end{align}
Hence, 
\begin{align}
	&\trace \left[\mathbb{E}  \left[ \widehat{\xv}_t\widehat{\xv}_t^T\right]\right]\\
	&= \trace \left[ \mathbb{E} \left[ \widehat{\Sigmam}  \widehat{\Sigmam}_\gamma  ^{-2} \left( \Sigmam^\frac{1}{2} \xv_t  + \deltav   \right)\left( \Sigmam^\frac{1}{2} \xv_t  + \deltav   \right)^T  \right] \right].
\end{align}
Observing that ${\xv_t}$, $\deltav$ and $\widehat{\Sigmam}$ are independent, and $\deltav \sim \mathcal{N} (\boldsymbol{0}, \frac{\Sigmam}{n})$,  we obtain
\begin{align}
	\trace \left[\mathbb{E}  \left[ \widehat{\xv}_t\widehat{\xv}_t^T\right]\right] 
	&= 	\trace \left[ \Sigmam \widehat{\Sigmam} \left( \widehat{\Sigmam}  + \gamma {\bf I}   \right)^{-2}  \right] + \dfrac{1}{n} 	\trace \left[ \Sigmam \widehat{\Sigmam} \left( \widehat{\Sigmam}  + \gamma {\bf I}   \right)^{-2}  \right] \\
	&=  \dfrac{n+1}{n}   \trace \left[ \Sigmam \widehat{\Sigmam} \left( \widehat{\Sigmam}  + \gamma {\bf I}   \right)^{-2}  \right].
\end{align}
Finally, the third term is obtained from
\begin{align}
	&\trace \left[\mathbb{E} \left[ \widehat{\xv}_t\xv_t^T\right]\right] \\
	&= \trace \left[ \mathbb{E} \left[ \widehat{\Sigmam}^\frac{1}{2} \widehat{\Sigmam}_\gamma ^{-1} (\yv - \widehat{\muv}) \xv_t^T \right]  \right]\\
	&= \trace \left[ \mathbb{E} \left[ \widehat{\Sigmam}^\frac{1}{2} \widehat{\Sigmam}_\gamma ^{-1}  \left( \Sigmam^\frac{1}{2} \xv_t  + \deltav   \right) \xv_t^T \right]  \right]\\	
	&= \trace \left[ \Sigmam^\frac{1}{2} \widehat{\Sigmam}^\frac{1}{2} \widehat{\Sigmam}_\gamma ^{-1}   \right].	 
\end{align}
\section{Proof of Theorem 1}
\label{sec:appC}
$\mathbb{E}[A(\gamma)]$ in \eqref{eqn:MSE1} can be directly obtained from \eqref{eqn:DE_2} with setting $\Thetam = \Sigmam$ and $z = -\gamma$. The second term, $B(\gamma)$, resulted from adding and subtracting $i\sqrt{\gamma} \Id_p$ with factoring $\widehat{\Sigmam}_\gamma^{-1}$  as follows:
\begin{align}
	\trace \Bigg[ \Sigmam^\frac{1}{2} \widehat{\Sigmam}^\frac{1}{2}  \widehat{\Sigmam}_\gamma^{-1} \Bigg]&=  \trace \Bigg\{    \Sigmam^\frac{1}{2}  \Bigg[  \Big ( \widehat{\Sigmam}^\frac{1}{2}  +i\sqrt{\gamma} \Id_p \Big)  -i\sqrt{\gamma} {\bf I} \Bigg] 
	. \Bigg [ \Big ( \widehat{\Sigmam}^\frac{1}{2}  +i\sqrt{\gamma} {\bf I} \Big) \Big (   \widehat{\Sigmam}^\frac{1}{2}  -i\sqrt{\gamma} \Id_p \Big) \Bigg ]^{-1}  \Bigg \}\\
	& = \trace \left[  \Sigmam^\frac{1}{2} \left ( \widehat{\Sigmam}^\frac{1}{2}  -i\sqrt{\gamma} {\bf I} \right)^{-1} \right]  -i\sqrt{\gamma}\trace\left [   \Sigmam^\frac{1}{2}  \widehat{\Sigmam}_\gamma ^{-1} \right] \label{eqn:thirdterm}.
\end{align}
We can further simplify \eqref{eqn:thirdterm} by noticing that the quantity on the left-hand side is a real quantity, so this implies
\begin{equation}
	\Im \left[   \trace \left[  \Sigmam^\frac{1}{2} \left ( \widehat{\Sigmam}^\frac{1}{2}  -i\sqrt{\gamma} {\bf I} \right)^{-1} \right] \right] = \sqrt{\gamma}\trace\left [   \Sigmam^\frac{1}{2}  \widehat{\Sigmam}_\gamma ^{-1} \right].
\end{equation}
Thus, we can express $B(\gamma)$ equivalently as 
\begin{equation}
	\trace \left[ \Sigmam^\frac{1}{2} \widehat{\Sigmam}^\frac{1}{2} \widehat{\Sigmam}_\gamma ^{-1}  \right] =\Re \left[   \trace \left[  \Sigmam^\frac{1}{2} \left ( \widehat{\Sigmam}^\frac{1}{2}  -i\sqrt{\gamma} {\bf I} \right)^{-1} \right] \right],
\end{equation}
so we can find $\mathbb{E} [B(\gamma)]$ easily from \eqref{eqn:DE_1} with setting $\Thetam = \Sigmam^\frac{1}{2}$, $z = i\sqrt{\gamma}$, and the third term in \eqref{eqn:MSE_DE} resulted.

\section{Proof of Theorem 2}
\label{sec:appD}
The MSE($\gamma$) expressed in \eqref{eqn:MSE_DE} converges to a sum of deterministic terms. To find a consistent estimator of \eqref{eqn:MSE_DE}, it is sufficient to  find a consistent estimator of each of these terms (Theorem 3.2.6 in \cite{1985-takeshi-advanced}).   

The consistent estimator, $\hat{\delta}_1$, can be derived using \eqref{eqn:delta_1} and \eqref{eqn:tilde_delta_1} in \eqref{eqn:DE_11}  ( with $\Thetam = \Id_p$),
\begin{align}
	\label{eqn:delta_hat_1}
	\dfrac{1}{n}\trace \left[  \widehat{\Sigmam}  \left( \widehat{\Sigmam} +\gamma \Id_p  \right)^{-1} \right] \asymp \dfrac{{\delta}_1}{1+{\delta}_1}.
\end{align}
The consistent estimator, $\hat{\delta}_1$, in \eqref{eqn:delta_1_hat} (that satisfies $\hat{\delta} \asymp \delta$) results immediately after rearranging \eqref{eqn:delta_hat_1}. The derivation of $\hat{\delta}_2$ follows similarly.

To derive the consistent estimator of $\phi \triangleq \dfrac{1}{n}\Bigl[ \Sigmam^2\bigl( \tilde{\delta}_1\Sigmam + \gamma {\bf I} \bigr)^{-2}   \Bigr] $ we differentiate $\delta_1$ to obtain
\begin{equation}
	\delta_1^{\prime} = \dfrac{\Psi}{1 - \phi(1+\delta_1)^{-2}},
\end{equation}
where $\Psi \triangleq \dfrac{1}{n}\trace [\Sigmam (\tilde{\delta_1}\Sigmam + \gamma\Id_p)^{-2}]$. Hence, we can estimate  ${\phi}$ consistently, (i.e., $\hat{\phi} \asymp \phi$) as
\begin{equation}
	\hat{\phi} = \dfrac{\hat{\delta}_1^\prime -\widehat{\Psi}}{\hat{\delta}_1^\prime(1+\hat{\delta}_1)^{-2}},
	\label{eqn:estPhi}
\end{equation}
where $\widehat{\Psi}$  is the consistent estimator of $\Psi$ can be estimated from \eqref{eqn:DE_22} when $\boldsymbol{\Theta} = {\bf I}$, $z=-\gamma$, as follows:
\begin{equation}
	\widehat{\Psi} = \dfrac{\dfrac{1}{n} 	\trace \left[ \widehat{\Sigmam} \left( \widehat{\Sigmam} +\gamma \Id_p  \right)^{-2} \right] }{(\hat{\tilde{\delta}}_1+\gamma\hat{\tilde{\delta}}_1^\prime)}.
\end{equation}
Substituting in (\ref{eqn:estPhi}), we can obtain the second term as 
\begin{equation}
	\dfrac{1}{n}\Bigl[ \Sigmam^2\bigl( \tilde{\delta}_1\Sigmam + \gamma {\bf I} \bigr)^{-2}   \Bigr]  
	\asymp \dfrac{(1+\hat{\delta}_1)^2}{\hat{\delta}_1^\prime } \Bigl[ \hat{\delta}^\prime_1 - \dfrac{\frac{1}{n} \trace \bigl[  \widehat{\Sigmam} \bigl(  \widehat{\Sigmam} + \gamma {\bf I}_p \bigr)^{-2}  \bigr]}{(\hat{\tilde{\delta}}_1+ \gamma\hat{\tilde{\delta}}_1^\prime)}  \Bigr].  
\end{equation}

\bibliographystyle{unsrt}  
\bibliographystyle{IEEEtran}
\bibliography{IEEEabrv,rmtbibfile}

%
%
%
%

\end{document}